\documentclass[12pt,preprint]{aastex}

\newcommand{\integral}{{\it INTEGRAL}}
\newcommand{\rxte}{{\it RXTE}}
\newcommand{\granat}{{\it GRANAT}}
\newcommand{\src}{{1E~1740.7--2942}}

\def\phs{ph~cm$^{-2}$~s$^{-1}$}
\def\chisq{$\chi^{2}$}
\def\rchisq{$\chi_{\nu}^{2}$}

\shorttitle{\src\ at high energy}
\shortauthors{Bouchet et al.}

\begin{document}

\title{Unveiling the high energy tail of 1E~1740.7--2942
with {\it INTEGRAL}\altaffilmark{*} }

\author{
L. Bouchet\altaffilmark{1}, M. Del Santo\altaffilmark{2}, E. Jourdain\altaffilmark{1}, J. P. Roques\altaffilmark{1}, A. Bazzano\altaffilmark{2}, 
G. De Cesare\altaffilmark{2,3,1}
        }
\altaffiltext{*}{INTEGRAL is an ESA project with instruments and science data centre
funded by ESA member states (especially the PI countries: Denmark, France, Germany, Italy, 
Spain, and Switzerland), Czech Republic and Poland with participation of Russia and USA.}
\altaffiltext{1}{CESR--Universite de Toulouse/CNRS, 9 Av. du Colonel Roche, 31028 Toulouse Cedex~04, France; jourdain@cesr.fr}
\altaffiltext{2}{INAF/Istituto di Astrofisica Spaziale e Fisica cosmica - Roma, via del Cavaliere 100, 00133 Roma, Italy} 
 \altaffiltext{3}{Dipartimento di Astronomia, Universita' degli Studi di Bologna, Via
Ranzani 1, I40127 Bologna, Italy}

\begin{abstract}
The microquasar \src\  is observed with \integral\ since Spring 2003. Here, we report on the source high energy
behaviour by using the first three years  of data collected with SPI and IBIS telescopes,
taking advantage of the instruments complementarity.
Light curves analysis showed 
two main states for \src: 
the canonical low/hard state of black-hole candidates and a ``dim'' state, characterised 
by a  $\sim$ 20 times fainter emission, detected only below 50 keV and when summing more than 1Ms of data.
For the first time the continuum of the low/hard state has been measured up to $\sim 600$ keV with a spectrum 
that is well represented by a thermal Comptonization plus an additional component necessary to fit the data above 200 keV. 
This high energy component could be related to non-thermal processes as already observed in other black-hole 
candidates.
Alternatively, we show that a model composed by two thermal Comptonizations provides an equally representative
description of the data: the temperature of the first population of electrons 
results as (kT$_{e}$)$_{1} \sim$ 30 keV while the second, (kT$_{e}$)$_{2}$, is fixed at 100 keV.

Finally, searching for 511 keV line showed no feature, either narrow or broad,
 transient or persistent. 
\end{abstract} 



\section{Introduction}
\src\ is a bright hard  X-ray source located at less than one 
degree off the Galactic Centre (Hertz \& Grindlay 1984; Cook et al. 1991; Roques et al. 1991),
classified as Black Hole Candidate (BHC; e.g. Sunyaev et al. 1991).
When Mirabel et al. (1992) discovered a double sided radio jet reaching large angular
distances from the core ($\sim$ 1'), the "microquasar" class was born with \src\
as its first member.
\\
All observations performed so far revealed that \src\ spends most of the time in the canonical
Low/Hard (LH) state of BHCs (Smith et al. 2002 and ref. therein). In this state, the
 X/$\gamma$ ray spectrum is empirically described 
by a power-law with a photon index of 1.4-1.5 plus a roll-over around 100 keV \cite{zdz00}. 
In a few occasions, soft spectral states have been observed during \src\ low flux levels \cite{smith02}.
Moreover, simultaneous \integral\ and \rxte\ broad-band spectral study
performed in 2003 report on an intermediate/soft spectral state
occurred just before the source quenching \cite{delsanto05}.

In 1990, the SIGMA telescope on-board \granat\ detected a broad line  
around the electron-positron annihilation energy (Bouchet et al. 1991; Sunyaev et al. 1991). This 
transient   feature appeared clearly during a 13 hours observation 
and then possibly in two further occasions but at a less significant level.  Numerous works dedicated to
similar line searches have followed and all led to negative conclusions (see for example Cheng et al. 1998 
and references therein). In this context, it is  interesting   
to perform a deep analysis of  SPI data in this energy domain,
and to seek for  any feature around 511 keV associated with \src.
\\
The superior energy resolution of the SPI telescope allows for a specific dedicated study 
of this topic. Indeed, for the first time, an instrument is capable to look for
a narrow feature in this particular source. During the first year  of observations, no evidence for point source
emission at 511 keV has been detected with SPI.  The upper limit at 3.5 $\sigma$ level is $1.6 \times 10^{-4}$ \phs\ for
a narrow line (Teegarden \& Watanabe 2006) while
the IBIS  data set a 2 $\sigma$ upper limit 
of $1.6 \times 10^{-4}$ \phs\ in the 535-585 keV energy band for 
an exposure time equal to 1.5 Ms (De Cesare et al. 2006).

We report here on the high-energy spectral properties as revealed with 
the \integral\ high-energy instruments.
The sensitivity and imaging capabilities of IBIS/ISGRI allow to determine the contribution of all 
the emitting sources in large fields of view, while the SPI telescope brings some 
additional spectral informations above 150-200 keV with a deep investigation of the 511 keV line status.

\section{\integral\ Observations}

Since its launch on October 17th 2002, \integral\ \cite{wink03} observed the Galactic
Center region two times per year, in the Spring and Fall visibility windows. 
Observations are performed in dither pattern with each pointing (named science window, SCW) lasting beetwen 1700 and 3600 seconds.
We have analysed all public data collected 
between Spring 2003 and Fall 2005 by the spectrometer
SPI \cite{vedrenne03} and the imager IBIS \cite{ube03}.

After image analysis and cleaning, the useful data set consists in about 3500 exposures
for a total useful time of 8 Ms divided in 6 periods (Spring and Fall, 2003, 2004 and 2005,
see Table \ref{log}).

\section{Data analysis}

\subsection{IBIS}

The unprecedented IBIS \cite{ube03} angular resolution combined with sensitivity ($<$1 mCrab
for 1 Ms ; Bird et al. 2007) 
allow us to resolve sources lying in crowded field, as the Galactic Centre. 
The IBIS Partially Coded Field Of View is 29$^\circ \times$29$^\circ$ 
at zero response, but the full instrument sensitivity 
is achieved in the 9$^\circ \times$9$^\circ$ Fully Coded Field of View.
For our aims, we selected  IBIS  observations 
including \src\  in the FOV up to 50\% coding (19$^\circ \times$19$^\circ$; see Tab. \ref{log}). 
In this paper, we refer to data collected with the IBIS low energy detector, ISGRI \cite{lebrun03}, 
covering the 15-1000 keV energy band.

The IBIS  scientific analysis has been performed using the
\integral\ off-line analysis software, OSA  \cite{goldwurm03}. 
The IBIS/ISGRI images have been extracted SCW by SCW in three energy bands, i.e. 
20-40~keV, 40-100~keV and 100-300~keV.
Mosaic images by revolution  have been used
to measure fluxes of all sources within 2 degree off \src\ used as input for SPI analysis (see
Section \ref{spibis}).
\\
Spectra have been extracted SCW by SCW in 35
logarithmic bins spanning from 20 keV to 600 keV.
The response matrices (RMF and ARF) used for spectral fitting 
are those delivered with OSA 5.1 distribution. To take into account
the improvements included in the matrices delivered in OSA-7, we modified
the ISGRI spectra by the factors corresponding to ratios between the 
Crab spectra measured respectively with OSA-5 and OSA-7 packages.

\subsection{SPI}
 In addition to its spectroscopic capability, SPI can 
image the sky with a spatial resolution of $2.6^\circ$ (FWHM ) over a field 
of view of $30^\circ$ (Roques et al. 2003). \\
The signal recorded by SPI camera consists of the contributions from sources in the field of view plus 
background. A system of equations is to be solved to determine sources and background intensities.
In order to reduce the number of unknowns necessary to describe the data, we introduce some known
 information on both components. For the background, 
the relative count rates of the 19 Ge detectors (uniformity maps) 
are very stable and can be kept constant within each considered period (see Table \ref{log}) 
while the global normalisation factor is determined by 6 hours intervals.
Concerning the sources, timescales are chosen in function of the source intensity and temporal behaviour, 
the faintest ones being  considered as constant.
Detailed description of the data analysis algorithms and methods, using matrices available in the OSA package,
 can be found in Bouchet et al. (2005; 2008).\\
Exposures were selected on the basis of their pointing direction 
which is here required to be less than $12^\circ$ from \src. 
This ensures to keep the maximum sensitivity for \src\ and reduces the
total field of view spanned by the observations, leading to
a simpler description of the sky.

\subsubsection{SPI correction from IBIS inputs}\label{spibis}
The region around \src\ is particularly crowded (Bird et al. 2007; Belanger et al. 2006).
Due to  the modest SPI angular resolution ($\sim 2.6^\circ$), the spectrum directly 
extracted at \src\ position may contain contributions from other weak/close/``not seen'' 
sources. Nevertheless, it is possible to obtain the emission spectrum of 
\src\ from SPI data thanks to the information provided by IBIS/ISGRI.\\

\begin{figure}
 \plotone{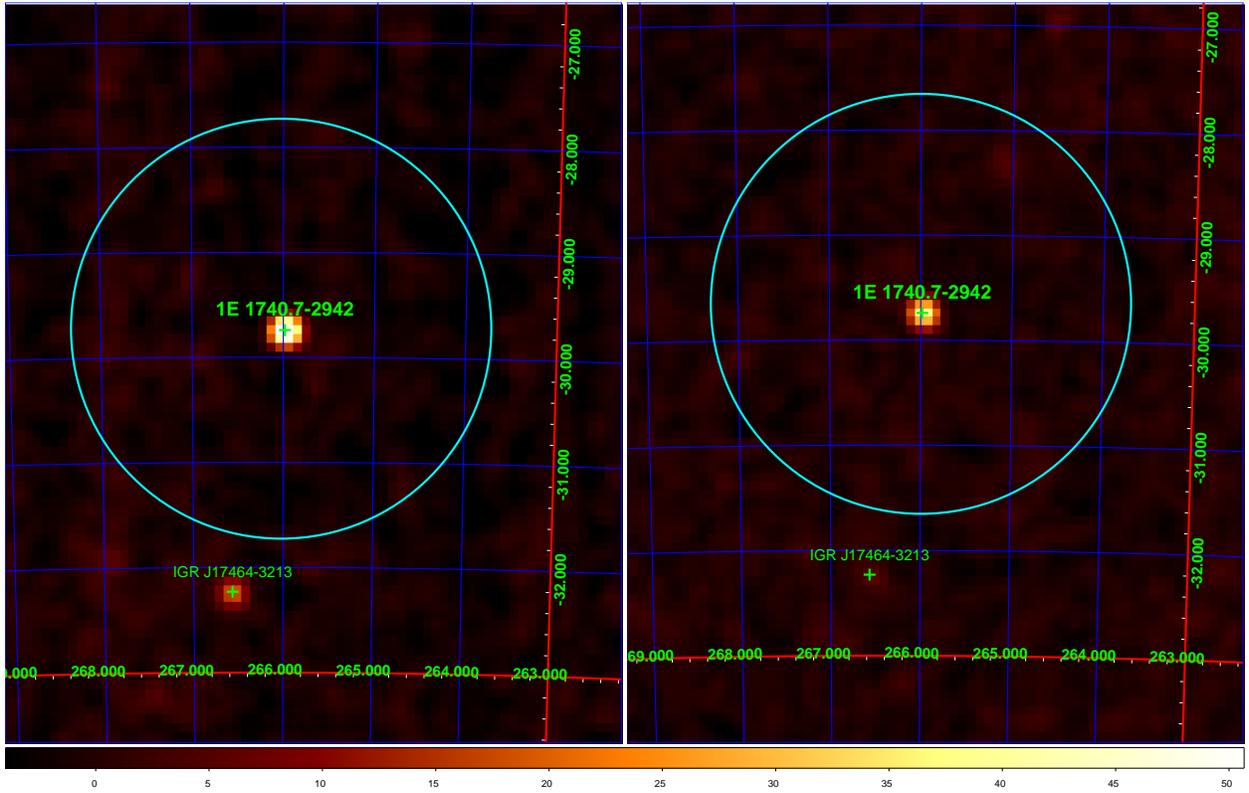} 
\caption{IBIS images of the 1E region in 2003 (left) and 2005 (right) in the 100-300 keV
energy range. The cyan circles are $2^\circ$ in radius.\label{image}} 
\end{figure}

For that, we need to determine the fraction of the flux extracted at the \src\ position
that  actually originates from the source itself. This has been determined by a set of simulations. 
The first step consists to extract the flux
of all emitting sources in its neighbourhood measured by IBIS. We then  
simulate the counts projected by them on the SPI detector plane, taking into 
account 
the complete (angular dependent) SPI response. Applying the 
standard analysis method to these simulated data  gives us a "\src\ region" flux, 
that we can compare to the \src\ flux injected as input in the simulation.
The ratio between these two fluxes corresponds to the factor we  
have to apply to correct the flux measured by SPI at the \src\ position in the 
observed data to obtain the flux attributable to the source itself.
This procedure has been repeated in a few broad bands and for each observational period.
This cleaning procedure takes into account a global contribution of all
potential contaminations, and the corresponding IBIS error bars can be considered as
very small  compared to the SPI ones. 
However, the contamination effect which is important in the low-energy  domain,
becomes negligible when going up to higher energies. In  fact, as can be seen 
in Fig. \ref{image}, above 100 keV only  \src\ is detected with  a significant flux within $2^\circ$  
off the source.

\subsubsection{Modelling the diffuse background : e+e- annihilation line and positronium emission}

The SPI design makes it sensitive to both source and diffuse emissions. On the other hand, 
the Galactic Centre (GC) region  is 
dominated in the $\sim$ 300 keV up to 
511 keV domain by the Galactic diffuse annihilation radiation. 
The annihilation line emission 
is detected with SPI with a flux of 
$\sim 1 \times 10^{-3}$ \phs\  
and a $\sim 8^\circ$  axisymmetric Gaussian spatial distribution   centered 
at the Galactic Centre \cite{kn05, bouchet08}. \src\ continuum emission around 511 keV
is expected to be (in the ``hard state'') 
of the order of a few percents of the 
galactic background line intensity. It is thus crucial for our work to determine this latter accurately.
This task has been performed using a larger data set (see Bouchet et al. 2008), which includes observations
 at larger latitudes and longitudes.
 
The  diffuse emission has been described in this process by two gaussians 
while eight known sources has been introduced as potential emitters (including \src).
Thus, the fitting algorithm (based on a \chisq\ minimization method, see Bouchet et al. 2008
for more details) is able to adjust simultaneously point sources fluxes and the diffuse component
contribution in the 511 keV line domain. The energy centroid and width of the positron annihilation
line were 
fixed at 511 keV and 2.5 keV FWHM respectively \cite{chura05}. The  fit procedure results in a
 model consisting in  two Gaussians with FWHM of 3.2$^\circ$  and 10.8$^\circ$  and fluxes of 2.3 and 7.0  
$\times 10^{-4}$ \phs\, respectively, as the best description of the 
annihilation line spatial distribution and flux, without any significant emission
from the point sources (see Weidenspointner et al.  2008,
 supplementary material, for independent analysis). \\
Concerning the positronium we assumed it to follow  the same two Gaussians
spatial distribution as the 511 keV line and determined its flux by the same fitting
procedure. This results in a positronium  fraction of 0.98, a value that is compatible with 
all SPI measurements \cite{bouchet08}.\\
Finally, the contributions of  these Galactic diffuse  components on the detector plane
are subtracted from the data in the counts space.\\

\section{Results}

\subsection{Temporal analysis}

Table \ref{flux} 
gives the \src\ averaged fluxes for the different periods in two broad bands.
The source mean  hardnesses in 2003 and 2005 are similar indicating that \src\ was in the 
LH state. Its intensity is rather stable on the revolution timescale, within 40 and 60 mCrab 
in the 20-40 keV energy band, except in the fall 2003 period, during which a continous decrease,
 from 85 to 27 mCrab, preceded the quenching observed in 2004 \cite{delsanto05}.
Indeed, in 2004, the source  was weaker with no detection above 5 $\sigma$ within individual revolutions. 
The data accumulation by periods allows to determine a mean flux of  a few mCrab.

\begin{figure}
 \plotone{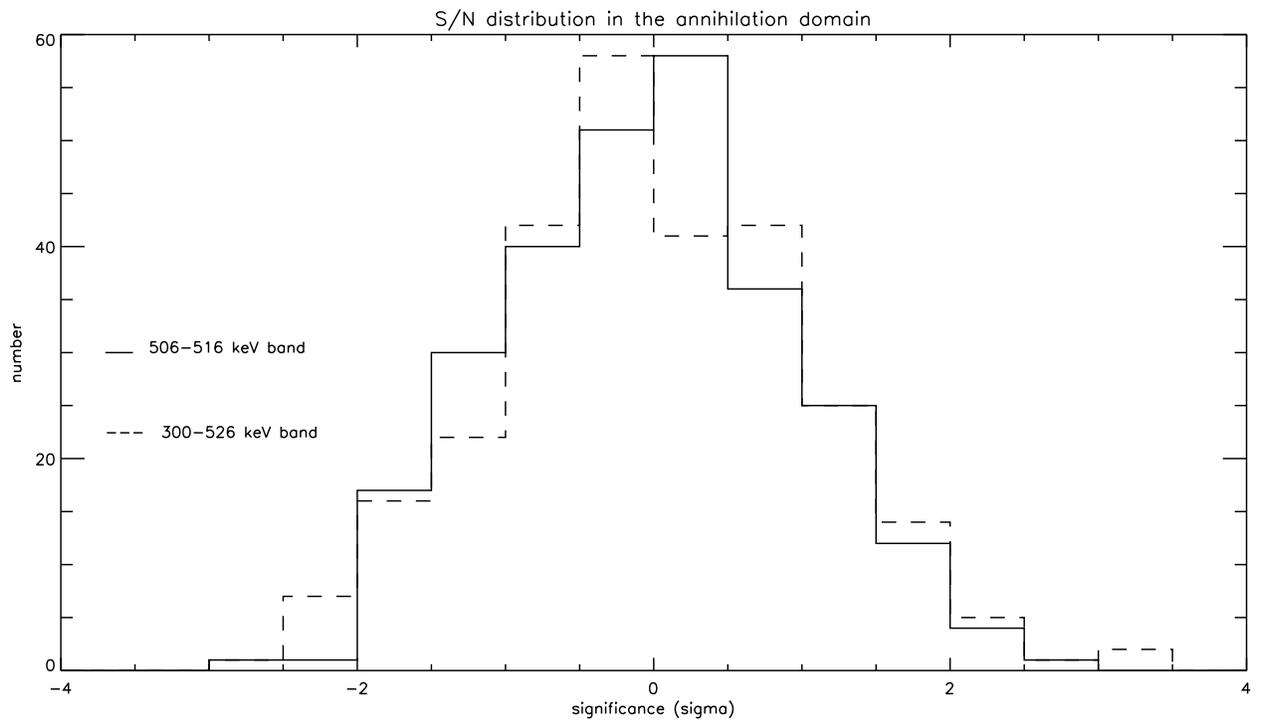} 
\caption{Distributions of the fluxes measured for the 511 keV line in sigma units, 
 for two hypotheses (a narrow feature: solid line, a broad feature: dashed line).\label{histo511}} 
\end{figure}

A dedicated study in a narrow (10 keV) band around  511 keV has been used to search for 
any transient emission from the annihilation process. We have tested 0.5 day  
and 1 day timescales  without detecting 
any significant emission. The actual
durations of each temporal bin  depends on the observational
planning and is thus variable. The 2 $\sigma$ upper limits range from $4.2 \times 10^{-4}$ \phs\ 
to $3.6 \times 10^{-3}$ \phs\ with an averaged value   $\sim 8 \times 10^{-4}$ \phs\, 
for a 0.5 day timescale, and  from $3.1 \times 10^{-4}$ \phs\ 
to $2.3 \times 10^{-3}$ \phs\ with an averaged value of $6.8 \times 10^{-4}$ \phs\,
for a day timescale. These results are illustrated by the distribution of the measurements in 
$\sigma$ unit for the 12 hours timescale (Fig \ref{histo511}, solid line) while
a 2 $\sigma$ upper limit of
$4.8 \times 10^{-5}$ \phs\ is deduced for the total
duration (see Table \ref{flux} for upper limits by periods).\\ 
Finally, a study has been performed for a broad feature, based on the 240 keV width (FWHM) reported in SIGMA data (Bouchet et al. 1991,
Sunyaev et al. 1991). The continuum emission (not negligible in such a broad band) 
 has been estimated by the mean flux over the considered period  and subtracted from the data. 
Here too, no  significant excess above the expected continuum emission  
can be claimed over
the 2003-2005 periods (Fig. \ref{histo511}, dashed line). 
The 2 $\sigma$ upper limits span from $1.7 \times 10^{-3}$ \phs\ 
to $1.3 \times 10^{-2}$  \phs\ with an averaged value close to $3 \times 10^{-3}$ \phs,  
for a 0.5 day timescale, and  from $1.2 \times 10^{-3}$ \phs\ 
to $8.8 \times 10^{-3}$ \phs\ with an averaged value of $2.5 \times 10^{-3}$ \phs, 
for a day timescale. Note that the line flux reported by SIGMA was $1.3 \times 10^{-2}$ \phs.
 
\subsection{Spectral analysis}
\begin{figure}
 \plotone{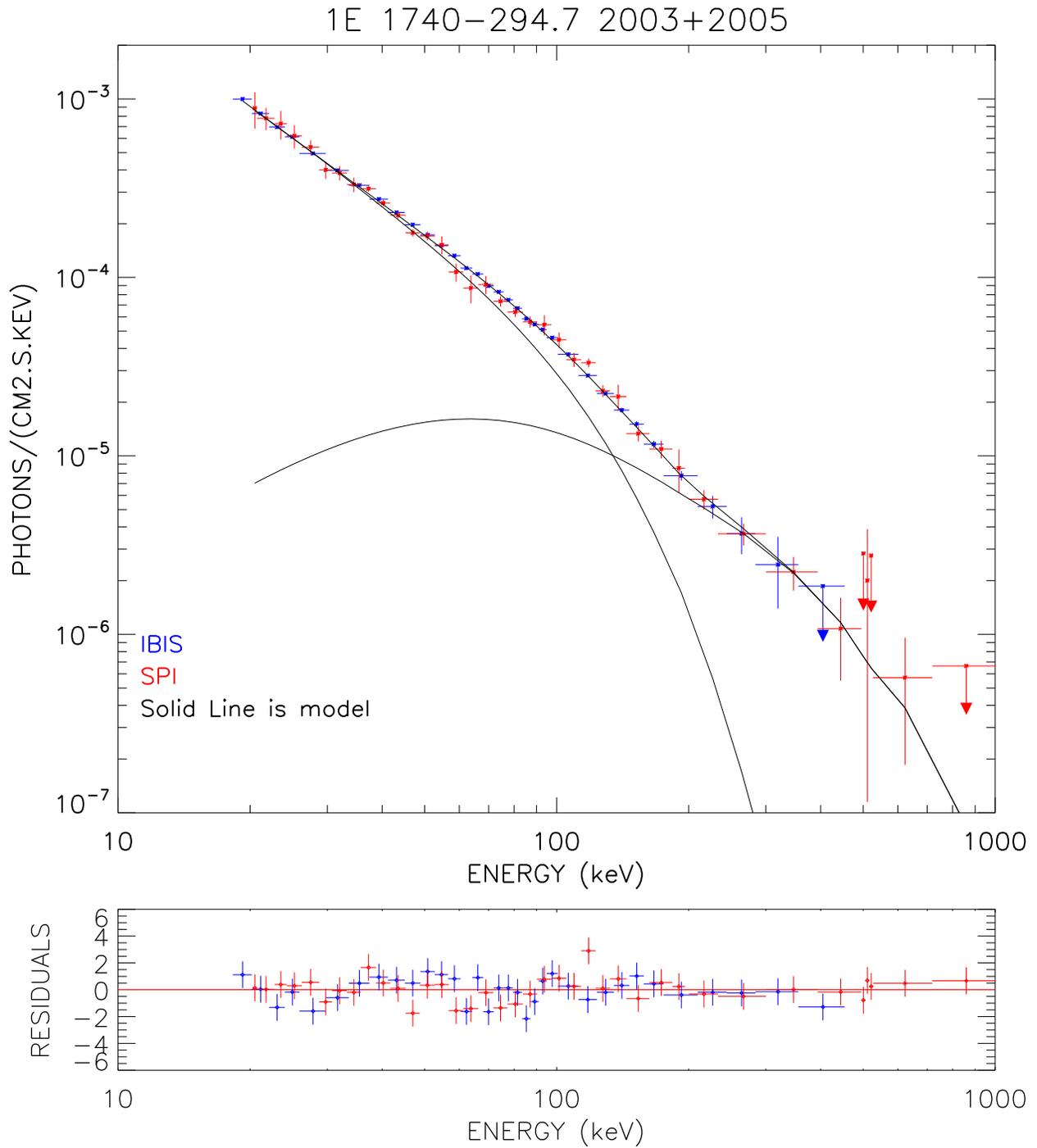}
\caption{1E~1740.7--2942 spectrum observed by IBIS and SPI for 2003 + 2005 observations.
Solid line is a two comptonisation model (2* comptt), with hotter temperature fixed to 100 keV.
The other parameters are  described in the text.\label{spectrum}} 
\end{figure}

After correction of the SPI data (Section \ref{spibis}), 
 SPI and IBIS/ISGRI spectra have been fitted simultaneously. 
We have first built averaged spectra for 2003 and 2005 separately. 
In a second step, being the source  in a similar state 
during these two periods, we achieved an averaged  LH state spectrum 
in order to obtain a better statistics at high energy.\\
Spectral fitting of these 3 data sets have been performed with the standard XSPEC v.11.3.1 tools.
We have included  a normalisation
factor during each fitting procedure and noticed that it remains between 0.94 and 1.0 
(ISGRI factor fixed to 1.0). Indeed, SPI spectra are 
very similar to the ISGRI ones (as illustrated in fig. \ref{spectrum}), even if they present  
some fluctuations at low energy, easily understandable in terms of
residual  cross-talk between neighbouring sources. However, this effect is limited 
and even  negligible above $\sim $ 100 keV.\\
During both periods, the source
emission extends up to 500 keV with a spectral shape presenting a clear cutoff around
$\sim$ 140 keV.
This cutoff is undoubtedly required by the \chisq\ statistics:
we obtain \rchisq\ of 10 and 6.6 (for 71 dof) with a power-law model, while 
adding a high-energy cutoff these values result as 0.9 and 1.14 (70 dof).
\\
Then we used a Comptonization model ({\sc{comptt}}, Titarchuk 1994) as this mechanism is expected 
to play the major role in our energy domain and to produce  such a cutoff.
We obtain electron temperatures (kT$_{e}$) of roughly 50 keV with optical depthes ($\tau$) close to 1 (see 
Tab. \ref{fit}), that are quite canonical values for this class of objects. 
However, the 2005 and total (2003+2005) spectra give 
high \rchisq\ values (1.35 and 1.8 for 69 dof). These, combined  with residuals at high energy,
suggest the presence of a supplementary 
component explaining data points above 200 keV. We have studied this hypothesis
in the total spectrum since its statistics allows us to better constrain 
the spectral parameters.
In order to model the high-energy data, we added a power law component  and obtain a 
\rchisq\ close to 1 with a photon index of 1.9 $\pm$ 0.1. The plasma temperature is thus 
decreased as  27 $\pm$ 2.2 keV with  $\tau$ = 1.9 $\pm$ 0.25.\\
Even though this two components model is quite satisfying, with a classical explanation of a non-thermal process
responsible of the power-law emission, as in High Soft States,  we also made an attempt to test
whether alternative scenario with only
thermal mecanisms were excluded. \\
Indeed,  a two (thermal) Comptonization model gives a similarly acceptable description
of the spectrum. 
The constraints on the parameters are very poor, so we impose a
 second population temperature at 100 keV and consider that it
comptonizes photons coming from the first Comptonizing region, i. e. (kT$_{seed}$)$_2$ = (kT$_{e}$)$_1 \sim$ 30 keV.
The optical depthes of both regions are found similar and compatible (1.6 $\pm$ 0.1 and 2.2 $\pm$ 0.8) 
while the \rchisq\ = 1.07 (67 dof) leads to an F-test probability of $\sim 10^{-8}$ for the existence of such 
a second component.  
Results corresponding to this scenario are displayed in Fig. \ref{spectrum} and  Tab. \ref{fit}. 
Since  this analysis is strongly based on ISGRI data at low energy, we performed a fit  using
SPI data above 100 keV only (no contamination effect) and found that the parameters are unchanged.

\section{Discussion and conclusions}

Even though the \src\  region is particularly difficult to analyse with the SPI telescope, we have demonstrated
that the use of the IBIS and SPI complementarity allows us to get a common spectrum of \src\  itself. Thanks
 to the inputs from the ISGRI detector, we have estimated the relative
contribution of all sources active in a $1^\circ$ circle around \src\  and showed that we can reconstruct
the  SPI \src\  spectrum with a precision better than 5 \% relatively to the ISGRI one.

Two main states have been observed: the canonical LH state with a 
 flux of $\sim$ 50  mCrab   and 60 mCrab  in the 20-40 keV and 40-100 keV bands respectively,
and a "dim" state during which the flux of \src\ is below the IBIS/ISGRI detection limit on the revolution timescale
and detected at a level of 2.6 -2.7 mCrab (14 $\sigma$) when integrated over $\sim$ 3 months periods.
\\
Spectra have been built for periods when the source was clearly detected (2003, 2005 and the sum of both).
The problematic of the SPI analysis leads to rather large error  bars but in all cases, the emission extends  
up to $\sim$ 500 keV, even though a high energy cutoff appears clearly in the data.
 When adjusted with a single Comptonization model,   an additional component is strongly required to fit
the data above 200 keV, particularly in the total spectrum because of the very significant emission
at these energies.\\  
This high energy component has been observed in several BHCs (e.g. McConnell et al. 2000; Zdziarski
et al. 2001), usually during high soft spectral states, and  explained
as  Compton up-scattering by a non-thermal electrons population \cite{zdz-gie04}. 
As alternative scenario, jets can easily produce hard X-ray emission via synchrotron radiation in addition to the inverse Compton scattering \cite{mark03}.
However, by computing radio-to-gamma ray Spectral Energy Distribution, Bosch-Ramon et al. (2006) 
ruled out the jet emission for the hard-X ray spectrum of \src,
favoring rather the corona origin. \\     
Recently, high energy excesses have been observed in transient BHCs when in LH state (i. e.  Del Santo et al. 2008; Joinet et al. 2007),
that could be the result of spatial/temporal variations in plasma parameters \cite{mal-jou00}.
We have demonstrated with our data that,  for the LH state spectrum of \src, a model consisting of two thermal
 Comptonization components, with a second hotter 
population ((kT$_{e}$)$_2$  fixed to 100 keV)  interacting with the photons produced by the first one,
  provides an interesting alternative to the non-thermal scenarios. This two 
   temperature model could  either correspond to  two distinct heating mecanisms/regions or
reflect the presence of a gradient of temperature in the Comptonising plasma.
 
Finally, even though the complexity of the considered region makes
 it difficult to attribute firmly the 
detected emission to \src, 
the presence of photons with energy greater than several hundreds of keV
 in a more or less persistent way (something as half or 2/3 of the time),
  together with previously reported annihilation emission,  support  a scenario in which 
\src\ is a source of positrons. Indeed, as proposed by van Oss \& Belyanin (1995),
 a plasma detected with a temperature much lower than 1 MeV is able to produce
positrons throught photon-photon absorption. The basic argument is that the 
hard X-ray emission comes from the regions close to the central black hole,
where the gravity field is very strong. The high local temperature is thus 
lowered, leading to an observed value far from  the relativistic domain, while pairs 
are created in the innermost disk and driven away. Annihilation outbursts could occur 
when the accretion flow intercepts the pair wind (van Oss \& Belyanin 1995). \\ 
Concerning the 511 keV line itself, no feature, broad or narrow, transient or persistent,
has been found, confirming the rare occurence of such a phenomenon in line with numerous
different studies already performed on this topics. It is worth to note, however,
 that SPI/INTEGRAL gives the first
opportunity to investigate it in terms of narrow feature (a few keV) with actual constraints on 
its parameters. 
Unfortunately, any strong conclusion would require more information on the 
expected duration (and width) of the emission.

\acknowledgements
We are grateful to ASI (contract I/080/07/0) and CNES for support.
As one of the thousands Italian researchers with a long-term temporary
position, MDS acknowledges the support of Nature (455, 835-836) and thanks
the Editors for increasing the international awareness of the current
critical situation of the Italian Research.

\clearpage

\begin{table} 
\caption{Observations log.\label{log}}
\begin{tabular}{ccccc}
\tableline\tableline
 Period & Start Time & End Time & Pointings (or Scw) & Exposure (Ms)    \\
        & (UT)        & UT       & IBIS  /  SPI            & IBIS  /  SPI \\           
\tableline
2003 spring  &  2003 Feb 28  &  2003 Apr 22  & 298 / 434  & 0.62 / 0.86  \\
2003 autumn    &  2003 Aug 19  &  2003 Oct 9   & 717 / 709 & 2.26 /  2.14  \\
2004 spring  &  2004 Feb 17  &  2004 Apr 20  & 547 / 544 & 1.29 / 1.18      \\
2004 autumn    &  2004 Aug 21  &  2004 Oct 28  & 747 / 610 & 1.78 / 1.35    \\
2005 spring  &  2005 Feb 16  &  2005 Apr 20  & 802 / 697 & 1.48 / 1.27    \\
2005 autumn    &  2005 Aug 16  &  2005 Oct 5   & 353 / 407 & 0.70 /  0.93    \\
\tableline

\end{tabular}
\end{table}

\begin{table} 
\scriptsize
\begin{center}
\caption{Averaged \src\ flux by periods. For the 511 keV feature, 2$\sigma$ upper limits are given 
for a narrow line ($\Delta $ E = 10 keV).\label{flux}}
\begin{tabular}{ccccccccccc}
\tableline\tableline
 Period &  2003 spring & 2003 fall  & 2004 spring & 2004 fall & 2005 spring &  2005 fall & 2003 \& 2005 & 2004   &   total period    &  \\
 \tableline
$F_{20-40 keV}$&51&75&2.6&2.7&53&53&62\\
 mCrab & & & & & & & & & &\\
\tableline
$F_{40-100 keV}$&60&95&-&-&56&54& 68 \\
 mCrab & & & & & & & & & &\\
\tableline 
 $F_{511keV}\times 10^{-4} $ &   $<$ 1.22  &  $<$ 0.66 &   $<$ 1.02 &  $<$ 1.10  & $<$ 1.10 &   $<$ 1.32 & $<$ 0.48 & $<$ 0.74 &  $<$ 0.40\\
 \phs\ & & & & & & & & & &\\
 \tableline
\end{tabular}
\end{center}
\end{table}

\begin{table}
 
\caption{The best fit  parameters for the combined ISGRI and SPI \src\ spectra with Comptonization
model ({\sc{comptt}}, Titarchuk 1994). T0 is fixed to 0.3 keV.    
For the 2003+2005 spectrum, the second line corresponds to a 2 Comptonization model ((kT$_{e}$)$_{1}$ , 
$\tau_{1}$, (kT$_{e}$)$_{2}$ , $\tau_{2}$) 
with the 2nd seed photon temperature fixed to (kT$_{e}$)$_{1}$  and (kT$_{e}$)$_{2}$  fixed to 100 keV (see text). \label{fit}}
\begin{tabular}{lcccccccccc}
\tableline\tableline
period & (kT$_{e}$)$_{1}$&$ \tau_{1}$ &  (kT$_{e}$)$_{2}$&$ \tau_{2}$ & \rchisq\ (dof) &   F-test $ ^{(1)}$\\
       &     keV    &    & keV     &                & &                 &          \\
\tableline 
2003 & 50.1 $\pm$ 2.6 &  1.00  $\pm$ 0.06 &&& 1.1 (69) & \\
2005 & 45.4 $\pm$ 2.7  & 1.10 $\pm$ 0.08  &&& 1.35 (69)& \\
2003+2005& 48.1 $\pm$ 2.0  & 1.06 $\pm$ 0.05 &&& 1.8 (69)&  \\
2003+2005& 29.4 $\pm$ 3.1 & 1.6 $\pm$0.1 & 100 & 2.2 $\pm$ 0.8 & 1.07 (67) & 10$^{-8}$  \\ 
\tableline
\end{tabular}
 \tablenotetext{(1)} { for the presence of a second Comptonization component.} 
  
\end{table}

\clearpage

 
\begin{thebibliography}{}
\bibitem[Belanger et al. 2006]{Belanger2006}
Belanger. G., Goldwurm, A., Renaud, M., et al. 2006, ApJ, 636, 275

\bibitem[Bird et al. 2007]{Bird07}
Bird, A.J., Malizia, A., Bazzano, A., et al. 2007, ApJS, 170, 175


\bibitem[Bosch-Ramon et al. 2006]{bosch06} 
Bosch-Ramon, V., Romero, G. E., Paredes, J. M., Bazzano, A., Del Santo, M.
\& Bassani, L. 2006, A\&A, 457, 1011

\bibitem[Bouchet et al. 1991]{Bouchet91} 
Bouchet, L.  et al. 1991, ApJ, 383, L45 


\bibitem[Bouchet et al. 2005]{Bouchet05}
Bouchet, L., Roques, J. P., Mandrou, P., Strong, A., Diehl, R., Lebrun, F. \& Terrier, R.
 2005, ApJ, 635, 1115


\bibitem[Bouchet et al. 2008]{bouchet08}
Bouchet, L., Jourdain, E., Roques, J. P., Strong, A., Diehl, R., Lebrun, F. \& Terrier, R. 2008, ApJ, 679, 1315

\bibitem[Cheng et al. 1998]{cheng98}
Cheng, L. X.,  Leventhal, M., Smith, D., Gehrels, N., Tueller, J. and Fishman, G., 1998, ApJ, 503, 809  

\bibitem[Churazov et al. 2005]{chura05}
Churazov, E., Sunyaev, R., Sazonov, S., Revnivtsev, M. \& Varshalovich, D.   2005, MNRAS, 357, 1377



\bibitem[Cook et al.\ 1991]{cook1991}
Cook, W.R., Grunsfeld, J. M., Heindl, W. A., Palmer, D. M., Prince, T. A.,
 Schindler, S. M. \& Stone, E. 1991, ApJL, 372, L75


\bibitem[De Cesare et al. 2006]{decesare06}
De Cesare, G., Bazzano, A., Capitanio, F., Del Santo, M., Lonjou, V., 
Natalucci, L., Ubertini, P. \& von Ballmoos, P. 2006, AdvSpR, 38, 1457

\bibitem[Del Santo et al. 2005]{delsanto05}
Del Santo, M.  et al. 2005, A\&A, 433, 613

\bibitem[Del Santo et al. 2008]{delsanto08}
Del Santo, M., Malzac, J., Jourdain, E., Belloni, T. \& Ubertini, P. 2008, 
MNRAS, in press. 
astro-ph 0807.1018

\bibitem[Goldwurm et al.\ 2003]{goldwurm03}
 Goldwurm, A. et al. 2003, A\&A, 411, L223


\bibitem[Hertz \& Grindlay 1984]{her84} Hertz, P. \& Grindlay, J. E. 1984, ApJ, 278, 137

\bibitem[Joinet et al. 2007]{joi07} 
Joinet, A., Jourdain, E., Malzac, J., Roques, J. P., Corbel, S., Rodriguez, J.,
 \& Kalemci, E. 2007, ApJ, 657, 400


\bibitem[Kn\"odlseder et al. 2005]{kn05}
Kn\"odlseder J. et ~al. 2005, A\&A, 441, 513
 
\bibitem[Lebrun et al. 2003]{lebrun03} 
Lebrun, F. et al. 2003, A\&A, 411, L141

\bibitem[Malzac \& Jourdain 2000]{mal-jou00} Malzac, J., \& Jourdain, E. 2000, A\&A, 359, 843

\bibitem[Markoff et al. 2003]{mark03}
Markoff, S., Nowak, M., Corbel, S., Fender, R., Falcke, H.  2003, A\&A, 397, 645

\bibitem[McConnell et al. 2000]{McCo00}
 McConnell, M. L.  et al. 2000, ApJ, 543, 928 

\bibitem[Mirabel et al. 1992]{Mirabel1992}
Mirabel, F., Rodriguez, L. F., Cordier, B., Paul, J. \& Lebrun, F. 1992, Nature, 358, 215


\bibitem[van Oss \& Belyanin 1995]{oss1995}
van Oss, R. F. \& Belyanin, A. A., 1991, A\&A., 302, 154

\bibitem[Roques et al. 1991]{Roques1991}
Roques, J.P.  et al.  1991, AdSpR, 11, 869

\bibitem[Roques(2003)]{Roques03}
Roques J.P.  et~al  2003, A\&A, 411, L91

\bibitem[Smith et al. 2002]{smith02} 
Smith, D. M., Heindl, W. A. \& Swank, J. H. 2002, ApJ, 569, 362


\bibitem[teegarden \& Watanabe 2006]{Teegar06}

Teegarden, B. J. \& Watanabe, 2006, ApJ, 646, 965

\bibitem[Sunyaev et al. 1991]{susu91} 
Sunyaev, R.  et al.  1991, ApJ, 383, L49

\bibitem[Titarchuk 1994]{tita94}
Titarchuk, L. 1994, ApJ, 434, 313


\bibitem[Ubertini et al. 2003]{ube03} 
Ubertini, P. et al. 2003, A\&A, 411, L131

\bibitem[Vedrenne et al. 2003]{vedrenne03}
Vedrenne, G. et al. 2003, A\&A, 411, L63


   
\bibitem[Weidenspointner et al. 2008]{weden08}
Weidenspointner, G. et al. 2008, Nature, 451, 159 


\bibitem[Winkler et al. 2003]{wink03}
Winkler, C. et al. 2003, A\&A, 411, L1
  
 \bibitem[Zdziarski 2000]{zdz00}
 Zdziarski, A. A. 2000, IAUS, 195, 153


 \bibitem[Zdziarski et al. 2001]{zdz01}
 Zdziarski, A. A., Grove, E., Poutanen, J., Rao, A. R. \& Vadawale, S. V.  2001, ApJL, 554, L45

 \bibitem[Zdziarski \& Gierli\'nski 2004]{zdz-gie04}
 Zdziarski, A. A., Gierli\'nski 2004, Progress of theoretical Physics, 155, 99

\end{thebibliography}
\end{document}